# Temperature Dependent Luminescent Decay Properties of CdTe Quantum Dot Monolayers: Impact of Concentration on Carrier Trapping.

*Graham P. Murphy‡, Xia Zhang‡, A. Louise Bradley\**

School of Physics and CRANN, Trinity College Dublin, College Green, Dublin 2, Ireland

ABSTRACT

We have investigated the temperature dependence of the photoluminescence spectra and average photoluminescence decay rate of CdTe quantum dot monolayers of different sizes as a function of concentration in the range 77 K to 296 K. It is shown that a simple three level analytic model involving bright and dark exciton states can only describe the lower temperature data but is unable to satisfactorily fit the data over the full temperature range. An extended model which includes external trap states is necessary to fit the data above approximately 150 K. Parameters for the model are obtained using both temporal and spectral data. The model indicates that the efficiency of interaction with trap states increases as the QD monolayer concentration increases, which is likely due to an increase in the density of available traps.

**Introduction**

Semiconductor quantum dots (QDs) are quasi-zero dimensional objects, typically of size 1-10 nm. QDs have interesting and unique optical properties which have seen them come to the fore in recent research replacing more traditional organic dyes and phosphors.[1–4] Highly tunable narrow emission, broad absorption, high quantum yields, with increased photostability[1,5,6] have led to QDs being used in wide variety of applications: such as bio-sensing[7] and bio-medical imaging,[8] photodetectors[9] and photovoltaics,[10,11] as well as LEDs.[4,12,13]

Previous studies investigating the temperature dependence of QDs have revealed the importance of considering the dark exciton state in order to explain the behavior of QD lifetimes at lower temperatures.[14,15] The dark state is long lived and weakly emitting due to spin transition rules. It lies below the nearest optically active state (bright exciton) in energy. At very low temperatures the thermal energy can be less than the bright-dark energy splitting, $k_{BT} < \Delta E$. This led Crooker *et al.* to postulate that excitons would be largely frozen into the dark state with no decay channels available, thus, providing an intrinsic upper limit to the exciton lifetime.[14] Since then there have been many studies investigating the temperature dependence of the photoluminescence properties of QDs for example; in solution,[16] in a polystyrene matrix,[17] embedded into a glass substrate,[18] as well as single core-shell QDs.[15] An analytic three-level model consisting of a ground state and the bright and dark states has been used extensively to fit experimental QD decay rates from as low as 2 K up to generally 140 K in a variety of QD systems.[15,16,19] At room temperature the fine structure of the exciton states are thermally mixed giving rise to an effective lifetime resulting from the mixing of the bright and dark states. Temperature dependent measurements in conjunction with the three-level model has allowed for the extraction of bright and dark exciton lifetimes, as well as the bright-dark energy splitting.

Bright and dark states in QDs were theoretically studied by Efros *et al.*[20] and the bright-dark splitting was found to be inversely dependent on size. Experimental studies investigating the size dependence of this splitting in various QDs[16,21–23] have revealed that the temperature dependence of QD lifetimes in the very low temperature range (< 20 K) is due to coupling between the dark exciton and acoustic phonon mode. Recently nano-engineering of the bright-dark splitting has become possible using advanced synthesis techniques in CdSe/CdS heterostructures. These include varying the number of CdS shells on a CdSe core[24] and changing the width of a nanorod shell of CdS around a spherical core of CdSe.[25] Increasing the activation energy of non-radiative processes such as Auger recombination has also been achieved in CdSe/CdS core/thick shell colloidal nanocrystals and could pave the way for 100 % quantum yield structures at room temperature.[26] For all the advances in CdSe based studies the literature is comparatively lacking in studies of water-soluble CdTe QDs which will be the focus of this work. Additionally, as mentioned above, much of the work to date has focused on temperatures below 150 K; this is because above this temperature non-radiative recombination becomes increasingly prevalent and thermal quenching of the PL intensity is non-negligible, leading to greater complexity. An extension to the three-level model that simultaneously considers both the exciton bright-dark splitting and carrier trapping was introduced to model temperature dependent decay rates of PbS QDs in the 10 K to 296 K range.[18]

We have investigated the temperature dependence of monolayers of two differently sized CdTe QDs. Monolayers of three QD concentrations have been studied for each size of QD. The monolayer concentration is varied from sparsely packed to densely packed to investigate the effect of concentration on the temperature dependent emission properties, spectral and dynamic. The PL decays of the QD ensemble in each monolayer is characterized by an average decay rate.

We will show that in order to accurately reproduce the temperature dependent data above 150 K it is essential to use a model which accounts for thermally activated carrier trapping in addition to the bright-dark splitting, and moreover that this trapping can become more efficient and more dominant at higher concentrations.

**Experimental Methods**

All structures were prepared using the layer-by-layer (LbL) assembly method which is based on sequential assembly of oppositely charged species onto a surface due to electrostatic forces.[27,28] Colloidal CdTe QDs stabilized with carboxylic acid (COOH) ligands were acquired commercially from PlasmaChem. The COOH ligand imparts a negative surface charge to the QDs. QDs of two different sizes were used in this study, hereafter referred to as QD1 and QD2. They have average diameters of (2.2 ± 0.1) nm and (3.8 ± 0.2) nm with peak emission wavelengths at room temperature of 557 nm and 690 nm respectively, as seen in Figure 1(a). QD diameter is calculated from the position of the first absorption peak.[29] The spectra shown in Figure 1(a) were measured for the highest concentration QD monolayer studied in each case, (3.4 ± 0.3) x $10^{17}$ m and (1.1 ± 0.3) x $10^{17}$ m$^{-2}$ respectively. A schematic of the structure is shown in Figure 1(b). Initially a quartz substrate is immersed in polyelectrolyte (PE) so that the PE can adsorb onto the surface of the quartz, subsequently four bilayers of PE are built up in order to provide a uniform surface for QD adhesion. The influence of the quartz vanishes after a few deposition cycles,[28] therefore the QDs will only be influenced by the polyelectrolyte surface and the QDs at each concentration experience the same substrate. The substrate is immersed in the QD solutions for varying times, thereby building up different concentrations of QDs in a monolayer. For QD1 the three concentrations studied were (1.2 ± 0.3) x $10^{17}$ m$^{-2}$, (2.7 ± 0.3) x $10^{17}$ m$^{-2}$, and (3.4 ± 0.3) x $10^{17}$ m$^{-2}$.

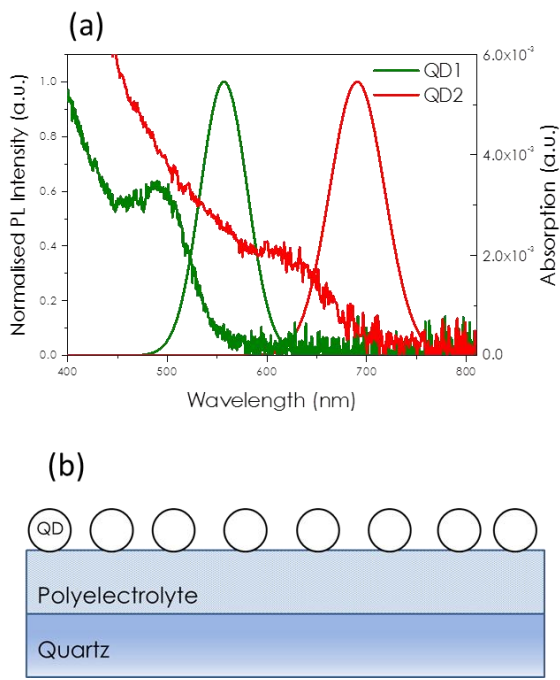

**Figure 1.** (a) Intensity normalized PL spectra (left axis) of QD1 (green lines) and QD2 (red lines) monolayers at room temperature (concentration c3). The absorption spectra (right axis) of the two QDs monolayers are also shown. (b) Schematic of the sample. QDs (white circles) are deposited onto PE layers that were firstly deposited on the quartz substrate to form a base layer for the QD monolayer deposition.

For QD2 they were $(1.9 \pm 0.3) \times 10^{16}$ m$^{-2}$, $(3.8 \pm 0.3) \times 10^{16}$ m$^{-2}$, and $(1.1 \pm 0.3) \times 10^{17}$ m$^{-2}$. These concentrations are denoted in order of increasing concentration c1, c2, and c3 for future reference. Absorption spectra are measured using a Perkin Elmer UV-vis spectrophotometer and are used in conjunction with the Lambert-Beer law to calculate the concentration of the QDs. Photoluminescence (PL) spectra are obtained from an Andor Shamrock sr-303i spectrometer with an Andor Newton 970EMCCD. This spectrometer is fiber coupled to an output port of a PicoQuant Microtime 200 Fluorescence Lifetime Imaging Microscope (FLIM) which is used to measure time resolved photoluminescence (TRPL), also known as the fluorescence lifetime. Having the PL spectrometer coupled to the FLIM system allows for the acquisition of TRPL and PL spectra from the same area of the sample with the same excitation source. The excitation density in this case was less than 1μJ/cm$^2$, to ensure single-exciton generation.[30] The samples were excited through a 40x long working distance objective using picosecond laser pulses at 405 nm with a repetition rate of 10 MHz, emission was collected back through the same objective. Lifetime decays were fitted with a two-exponential decay function

$$I(t) = A_1 exp[-t/\tau_1] + A_2 exp[-t/\tau_2] \quad (1)$$

where $A_1$ and $A_2$ are the intensity amplitudes of the two decays with lifetimes $\tau_1$ and $\tau_2$, respectively. The average lifetime $\tau_{av}$ is then calculated from an intensity weighted mean

$$\tau_{av} = \frac{A_1 \tau_1^2 + A_2 \tau_2^2}{A_1 \tau_1 + A_2 \tau_2} \quad (2)$$

The average decay rate is then easily determined from this lifetime; $k_{av} = 1/\tau_{av}$. A Janis ST-500 liquid nitrogen continuous flow cryostat was integrated into the FLIM system enabling measurement of TRPL and PL from ~ 77 K up to room temperature (~ 296 K).

## Results and Discussion

**Steady-State Photoluminescence**

The PL spectra show strong temperature dependent properties, shown in Figures 2 and 3. As the temperature increases the PL intensity decreases, as seen in Figure 2, and this trend is observed over the entire measurement range with no luminescence temperature anti-quenching observed.[31]

In addition to the PL intensity decrease, the emission energy red shifts, and the full-width at half-maximum (FWHM) of the spectra increases with increasing temperature.

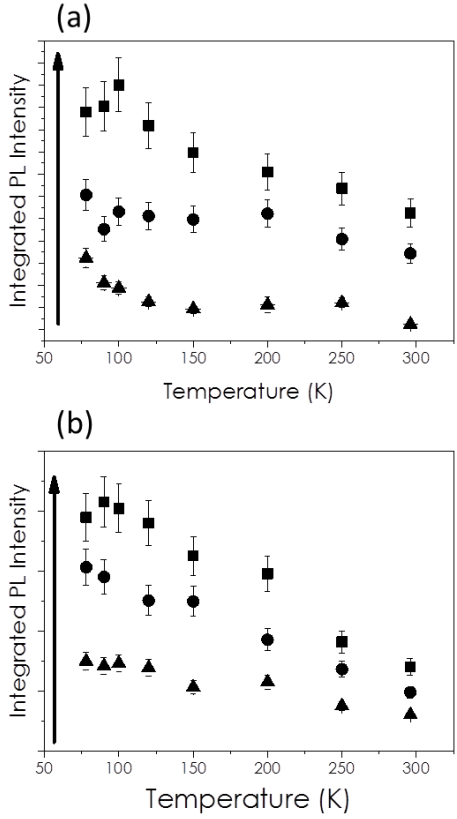

**Figure 2**: a[(b)] Integrated PL for QD1[QD2] as a function of temperature. The arrows indicate increasing concentration, triangles for c1, circles for c2, and squares for c3.

The common approach for fitting the peak shift and FWHM is to use expressions originally used to describe the temperature dependence of the band-gap and excitonic peak broadening in bulk semiconductors.[17,18,32] The temperature dependence of the PL broadening is analyzed with the following[33]

$$FWHM(T) = \Gamma_{inh} + \sigma T + \Gamma_{LO}[exp(E_{LO}/k_B T) - 1]^{-1} \qquad (3)$$

where $\Gamma_{inh}$ is the temperature independent inhomogeneous broadening (due to QD size, shape, etc.), $\sigma$ is an exciton-acoustic phonon coupling constant, and $\Gamma_{LO}$ represents the exciton-LO phonons coupling coefficient.

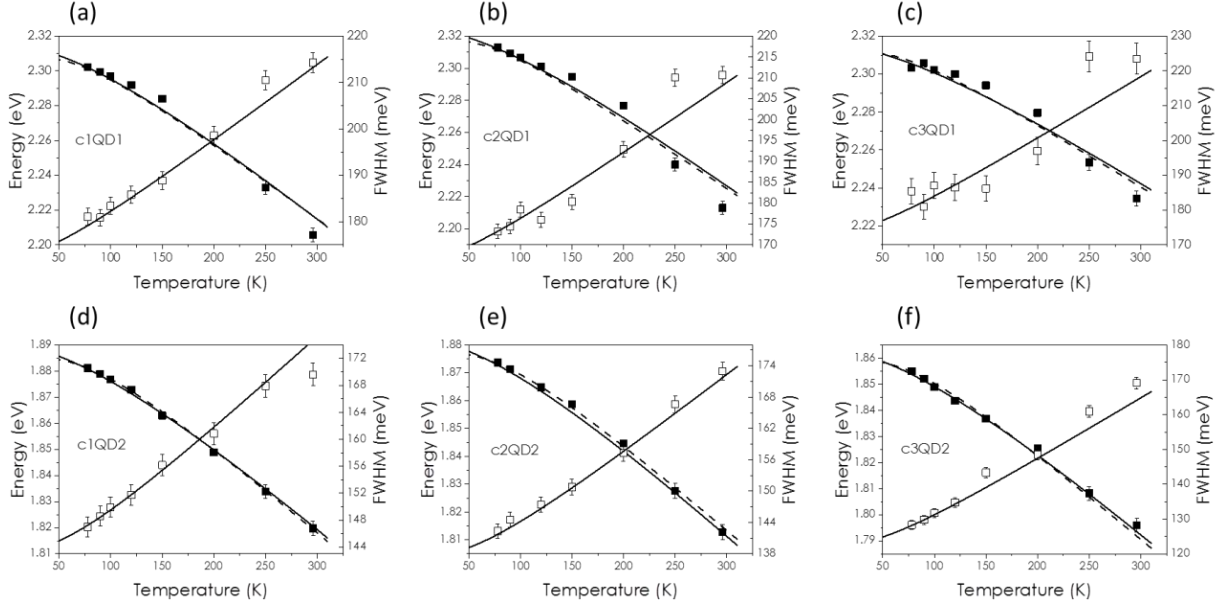

**Figure 3.** (a) – (c)[(d) – (f)] PL peak energy (full squares, left axis) and FWHM (empty squares, right axis) as a function of temperature for QD1[QD2] with increasing concentration from left to right. For example, label c1QD1 refers to the lowest concentration, c1, for QD1. For the peak PL energy the solid lines and dashes are best-fit curves for Equations (4) and (5) respectively. For the FWHM the solid lines are best fit curves for Equation (3).

It is found that the FWHM can be fitted by Equation (3) using a single value of $E_{LO}$ for each QD, $E_{LO} = (18 \pm 5)$ meV and $E_{LO} = (20 \pm 3)$ meV for QD1 and QD2, respectively. This suggests that the temperature dependence is dominated by the influence of carrier trapping. The other fitting parameters are found to be $\sigma = (90 \pm 10)$ μeV/K for QD1 and $\sigma = (60 \pm 10)$ μeV/K for QD2. This value of $\sigma$ is much larger than the estimated value for bulk CdTe due to quantum confinement.[17,32] The increase of $\sigma$ with decreasing QD size is in qualitative agreement with the increase of acoustic phonon coupling with increasing confinement in CdTe quantum wells.[34] Similarly, the value for $\Gamma_{LO}$ found for QD2 ($\Gamma_{LO}$ = (20 ± 3) meV) is in agreement with that reported by Morello,[32] and for smaller QD1 we find $\Gamma_{LO}$ = (18 ± 5) meV, which agrees within error. These values are smaller than the theoretical bulk value (24.5 meV) due to quantum confinement.[32,35]

The experimental data for the peak position of the PL emission can be fitted with the Varshni relation[36]

$$E_g(T) = E_g(0) - \alpha \frac{T^2}{(T + \beta)} \tag{4}$$

where $E_g(0)$ is the energy gap at 0 K, $\alpha$ is the temperature coefficient, the value of $\beta$ is close to the Debye temperature of the material. To fit we keep $\beta$ constant at the bulk value of 158 K and find best fit values for $\alpha$, yielding $\alpha$ = (4.6 ± 0.8) x $10^{-4}$ eV/K and $\alpha$ = (3.5 ± 0.2) x $10^{-4}$ eV/K for QD1 and QD2, respectively. The $\alpha$ value for QD2 is close to the bulk value, 3 x $10^{-4}$ eV/K, and also consistent with previously reported values for CdTe QDs.[32] QD1 is slightly larger with greater uncertainty, however, this is still relatively close to the bulk value.

To further confirm the average value of $E_{LO}$ the PL emission peak as a function of temperature can be fitted with a second equation first proposed by O'Donnell and Chen[37]

$$E_g(T) = E_g(0) - S\langle E_{LO}\rangle \left[\coth\left(\frac{\langle E_{LO}\rangle}{2k_BT}\right) - 1\right] \tag{5}$$

in which $S$ is a dimensionless coupling constant known as the Huang-Rhys parameter and $\langle E_{LO}\rangle$ is the average phonon energy. In this case fits can be obtained for QD1 with $\langle E_{LO}\rangle$ = (18 ± 5) meV and $S$ = 2.6 ± 0.3 and for QD2 with $\langle E_{LO}\rangle$ = (20 ± 3) meV and $S$ = 1.9 ± 0.1, which is once again in agreement with what is obtained from fitting the FWHM data. The decreasing value of $S$ with increasing size is in qualitative agreement with results presented previously on CdTe.[38] Since the values of $E_{LO}$ are within error for the two QDs all subsequent fitting using this parameter will be done with a single value of 20 meV.

It is interesting to note that the temperature dependent properties of the ensemble PL spectra could be fit without any consideration of a temperature dependent intra-energy transfer or exciton migration between the QDs within a monolayer, even for the highest concentration monolayers. This suggests that it is a not a dominant factor in the temperature dependence of the spectral properties of an individual QD monolayer and that it is not manifesting as a strong effect on the temperature dependence as the QD concentration is increased. Intra-energy transfer within QD ensembles, including QD monolayers, has been well-studied.[30,39,40] A spectral feature of intra-energy transfer is red-shifting of the PL spectrum with increasing QD concentration. Only a relatively small red-shift is observed as the QD concentration was increased, Figure 3 (a-c) and (d-f) for QD1 and QD2, respectively. Intra-energy transfer also manifests in spectrally-resolved TRPL measurements, showing a shortening of the photoluminescence decay rate on the high energy side of the spectrum, with a corresponding increase on the low energy side of the spectrum. However, in this study the time-resolved PL measurements are not spectrally-resolved and the measured decay rate represents an average for the entire ensemble.

The temperature dependence of the PL intensity due to the onset of thermally activated carrier trapping can be described by[32]

$$I_{PL}(T) = \frac{I_0}{1 + a[exp(-E_a/k_B T)] + b[exp(E_{LO}/k_B T) - 1]^{-q}} \tag{6}$$

where $I_0$ is the 0 K integrated PL intensity, $q$ is the number of LO phonons involved in thermal escape of carriers, $E_{LO}$ is their energy, $k_B$ is Boltzmann's constant and $E_a$ is an activation energy. The trends of the PL intensity at temperatures lower than 77 K can give insight into the nature of the activation energy. It has been attributed to thermally activated transfer from the dark exciton state to the bright state[14] when coupled with a blue shift and an increase in the intensity. Many other reports have found a value of $E_a$ much larger than the bright-dark splitting energy and have thus attributed it to trapping at surface/interface defect states.[17,32,38] Such attributions are beyond the scope of this work, as we focus on the effect of carrier trapping at temperatures above 150 K. It can be noted however, that using the value of $E_{LO}$ extracted from the FWHM and peak energy fitting, in conjunction with a value of $E_a$ corresponding to the bright-dark splitting energy (determined later in the paper by fitting the decay rates), it is possible to fit the measured data for the PL intensity. However, as the measurements herein are taken down to approximately 77 K and not towards 0 K there is a large uncertainty in the value of $I_0$, which can strongly influence the other fit parameters. Therefore, we have focused on the FWHM and peak energy fitting for extracting the appropriate parameters, in particular, $E_{LO}$.

**Fluorescent Lifetimes and Decay Rates**

In this section we will consider the temperature dependence of the QD lifetimes and decay rates. Using an analytical model it will be shown that is important to take account of both bright and dark states and at high temperatures (> 150 K) it is vital to include the effects of carrier trapping. Figure 4 shows PL decays for the highest QD concentration monolayers for each QD, c3QD1 (shown in green) and c3QD2 (shown in red), at 296 K (lines) and 77 K (dots). It can be seen that the PL lifetime increases at lower temperatures, as is well documented.[14–16,19,23,41] The average lifetime is determined using Equation (2), yielding $\tau(296\ K)_{QD1} = (6.0 \pm 0.2)$ ns, $\tau(77\ K)_{QD1} = (13.5 \pm 0.2)$ ns, $\tau(296\ K)_{QD2} = (14.4 \pm 0.5)$ ns and $\tau(77\ K)_{QD2} = (20.9 \pm 0.5)$ ns for the examples shown. The average decay rate, $k_{QD} = \tau_{QD}^{-1}$, can be determined from the PL decays at each temperature. Figures 5 (a) and (b) show the temperature dependence of the average decay rate for monolayers of both QD for each of the three QD concentrations. As expected the decay rate increases as the temperature increases over the entire measured range for both QDs. It is also noted that the overall decay rate increases with increasing concentration and the decay rate increases more sharply at higher temperatures for the higher concentrations.

Analytic expressions for the decay rates as a function of temperature can be derived using a model consisting of a ground state, $|g\rangle$, and two excited states: a lower energy dark exciton state, $|Dx\rangle$, and a higher energy bright exciton state, $|Br\rangle$.[14,15,19,20]

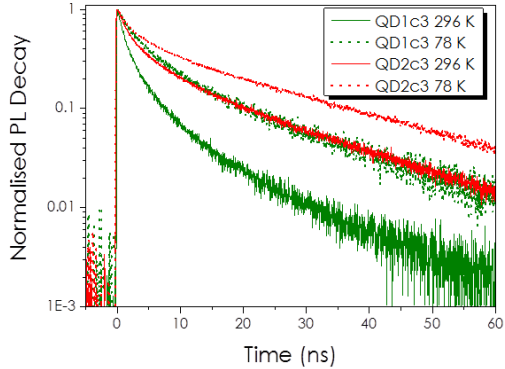

**Figure 4**. Normalised PL decays at room (~296 K) and low (~77 K) temperature for QD1 (green lines and dots, respectively) and QD2 (red lines and dots, respectively). The PL shown are from the highest QD concentration monolayers for both QDs, denoted c3. The average lifetime obtained from fitting the decays are $\tau(296\ K)_{QD1} = (6.0 \pm 0.2)$ ns, $\tau(77\ K)_{QD1} = (13.5 \pm 0.2)$ ns. $\tau(296\ K)_{QD2} = (14.4 \pm 0.5)$ ns and $\tau(77\ K)_{QD2} = (20.9 \pm 0.5)$ ns.

A schematic of this model is shown as the inset of Figure 5 (a). The dark state is long lived and weakly emitting due to spin transition rules, whereas the bright state is short lived and strongly emitting. The energy splitting, $\Delta E$, between the bright and dark states is generally in the range of a few meV, depending on the size, shape, and type of the QDs.[16,21–23] The rate equations for the number of excitons in the model is given as[19]

$$\frac{dN_{Br}}{dt} = -k_{Br}N_{Br} - k_{rel}N_{Br} + k_{rel}exp(-\Delta E/k_B T)N_{Dx}$$

$$\frac{dN_{Dx}}{dt} = -k_{Dx}N_{Dx} + k_{rel}N_{Br} - k_{rel}exp(-\Delta E/k_B T)N_{Dx} \qquad (7)$$

where $N_{Br}$ ($N_{Dx}$) and $k_{Br}$ ($k_{Dx}$) give the population and the radiative decay rate for the bright state (dark state), respectively and $k_{rel}$ is the relaxation rate from the bright state to the dark state. Relaxation rates between bright and dark states have been found to be of the order of fractions of ps$^{-1}$ for CdTe QDs,[42] which is orders of magnitudes faster than the PL decay times studied here. Combining these two, the rate equation for the total number of excitons, N is given by

$$\frac{dN}{dt} = -(k_{Br})N_{Br} - (k_{Dx})N_{Dx} = -Nk_{QD}(T) \qquad (8)$$

where $k_{QD}(T)$ is the decay rate of the QD PL. The population of the bright and dark states gives the total number of excitons ($N_{Br} + N_{Dx} = N$). It can be seen that the relaxation terms cancel in the expression for $\frac{dN}{dt}$. Assuming a Boltzmann distribution of excitons between |Br⟩ and |Dx⟩ on the basis of a statistical ensemble of QDs[14,18,19] then $N_{Br}/N_{Dx} = exp(-\Delta E/k_B T)$. Solving Equation (8) yields

$$k_{QD}(T) = \frac{k_{Dx} + (k_{Br})\, exp[-\Delta E/k_B T]}{1 + exp[-\Delta E/k_B T]} \tag{9}$$

The form of the temperature dependence of the decay rate can be compared to the experimental data in Figure 5. The fitting parameters are shown in Table 1, the value of ΔE extracted from these fits is of the order of those previously reported[21,22] and it also decreases with increasing QD size which again is in agreement with previous reports.[16,21–23] The value of $k_{Br}$ is seen to increase with increasing concentration for both QDs while the value of $k_{Dx}$ is relatively static. This would be in agreement with the assumption that the dark exciton rate gives a limit to the overall value of the decay rate at low temperature. It is obvious, however, particularly for QD1, that this model is incomplete as it does not fit the data well at higher temperatures, especially for the higher QD concentrations where the discrepancy is largest.

The above model does not consider external traps which are known to strongly influence the PL emission of QDs at higher temperatures.[14,16–18,32,43] If the discrepancy between the model and data is accounted for by this additional non-radiative recombination path, then the data indicates that at higher temperatures non-radiative recombination and carrier trapping is much more dominant for samples with higher concentrations. Qualitatively this could be attributed to an increase in the density of available traps, where either a single QD can effectively "see" defects or traps associated with other QDs nearby or the close proximity of QDs may lead to the formation of additional defects. With this in mind comparing the two QDs is interesting as the lowest concentration of QD2 fits very well with Equation (9) over the whole temperature range while there is still significant discrepancy for QD1.

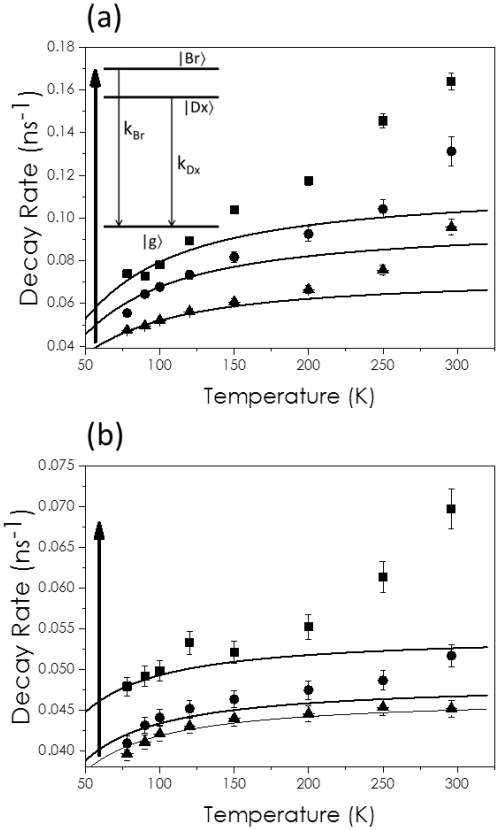

**Figure 5.** Temperature dependence of decay rate for monolayers of varying QD concentration c1-c3 (a) QD1 and (b) QD2. The arrows indicate the direction of increasing concentration. The solid lines are fits using Equation (9). The fitting parameters are presented in Table 1. Inset: Schematic of model.

The defect concentration has a strong influence on the optical properties of nanocrystals and smaller QDs typically exhibit lower quantum yields.[44] The smaller QD1 has a relatively higher surface to volume ratio. There is a higher concentration of defects at the surface, giving rise to a higher density of mid-gap states acting as carrier traps. In addition, the lowest concentration studied for QD1 is approximately 6 times greater than the lowest concentration for QD2. Therefore, despite the fact that the diameter of QD2 is significantly larger than QD1 the average separation of QDs for the lowest concentration QD2 monolayer is greater. As discussed above this could contribute to a lower density of defects being available for non-radiative recombination for this sample.

**Table 1**. Fitting parameters for decay rates based on Equation (9).

|  | $k_{Br}$ | $k_{Dx}$ | $\Delta E$ |
|---|---|---|---|
| **QD1** | ±4% | ±10 % | ±7% |
| c1 | 0.128 ns$^{-1}$ | 0.019 ns$^{-1}$ | 7 meV |
| c2 | 0.175 ns$^{-1}$ | 0.020 ns$^{-1}$ | 7 meV |
| c3 | 0.208 ns$^{-1}$ | 0.022 ns$^{-1}$ | 7 meV |
| **QD2** | ±3% | ±6% | ±13% |
| c1 | 0.068 ns$^{-1}$ | 0.025 ns$^{-1}$ | 4 meV |
| c2 | 0.070 ns$^{-1}$ | 0.026 ns$^{-1}$ | 4 meV |
| c3 | 0.076 ns$^{-1}$ | 0.032 ns$^{-1}$ | 4 meV |

To consider the role of such trapping in more detail the basic three level model introduced earlier can be extended by introducing an external trap state, $|T\rangle$, in addition to the bright and dark states, see the inset of Figure 6 (a) for schematic. In this case the rate equation becomes

$$\frac{dN}{dt} = -(k_{Br} + \kappa_{Br,n})N_{Br} - (k_{Dx} + \kappa_{Dx,m})N_{Dx} = -Nk_{QD}(T) \quad (10)$$

where $\kappa_{Br,n}$ and $\kappa_{Dx,m}$ are non-radiative relaxation from the bright and dark states to the trap states involving an absorption of *n* and *m* phonons with an energy $E_{ph}$. These can be expressed as[18]

$$\kappa_{Br,n} = k_0[exp(E_{ph}/k_BT) - 1]^{-n} \quad (11)$$

$$\kappa_{Dx,m} = k_0[exp(E_{ph}/k_BT) - 1]^{-m} \quad (12)$$

where $k_0$ is a rate constant characterizing the efficiency of the thermally induced non-radiative relaxation to trap states, and $m$ and $n$ refer to the number of phonons required for carrier escape from the dark and bright state, respectively. Solving Equation (10) leads to

$$k_{QD}(T) = \frac{k_{Dx} + \kappa_{Dx,m} + (k_{Br} + \kappa_{Br,n})exp[-\Delta E/k_BT]}{1 + exp[-\Delta E/k_BT]} \quad (13)$$

and substituting in Equations (11) and (12) leads to

$$k_{QD}(T) = \frac{ak_{Br} + k_{Dx} + k_0[a(b-1)^{-n} + (b-1)^{-m}]}{1 + a} \quad (14)$$

with $a = exp[-\Delta E/k_BT]$, $b = exp[E_{ph}/k_BT]$. We can attribute the phonon energy in this case, $E_{ph}$ to the average LO phonon energy of 20 meV calculated from the analysis of the PL spectral properties discussed earlier. The other initial parameters are taken from the earlier fitting using Equation (9), which was able to fit the low temperature data for both QDs at all concentrations. All parameters are shown in Table 2. The parameters $k_{Br}$, $k_{Dx}$, and $\Delta E$, taken from Table 1, are held fixed in this fitting, and therefore $k_0$, $m$, and $n$ are the fitting parameters. As can be seen in Figure 6, incorporating thermally activated carrier trapping allows us to reproduce the data quite nicely. This shows that the decay rates which prove accurate for fitting at the low end of the temperature range remain unchanged when introducing thermally activated carrier trapping, and that including this trapping accounts for all of the discrepancy between the data and the model presented in Equation (9) at the higher temperatures. This further indicates that thermally activated carrier trapping is the dominant feature at higher temperatures. From these fit parameters the trap state is found to be 37 meV and 52 meV above the bright state for QD1 and QD2, respectively. A wide range of activation energies for thermal trapping has been reported for CdTe nanocrystals. The discrepancies have been attributed to variations in surface passivation, with higher values generally associated with lower defect densities.[32,45,46] It can be noted that the values reported here are close to the activation energy of 46 meV reported by Chon *et al*. for 4.6 nm CdTe QDs.[45]

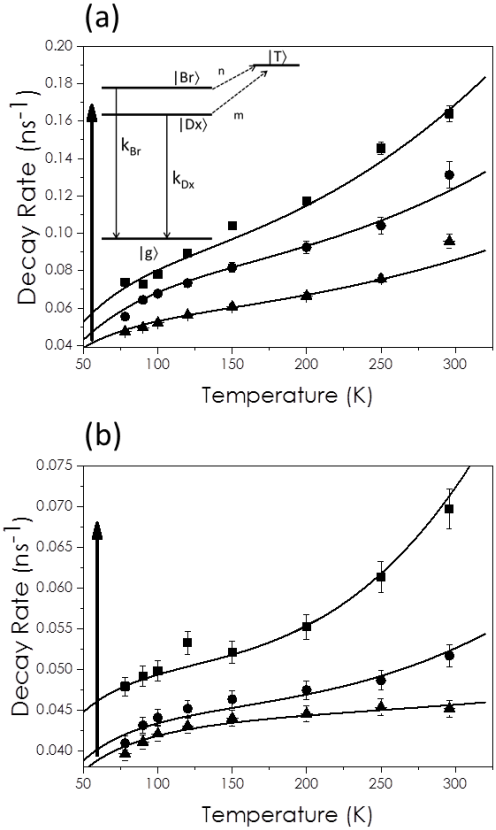

**Figure 6.** Temperature dependence of decay rates for (a) QD1 and (b) QD2 with increasing concentrations. Solid lines are fits to Equation (14). See Table 2 for fitting parameters. Inset: Schematic of model including trap state.

Focusing on the efficiency of the thermally induced non-radiative relaxation to trap states reflected in the rate $k_0$, it can be seen that it is larger for QD1 and for both QDs it increases as QD concentration increases. The larger $k_0$ for QD1 is in agreement with the higher defect density expected for smaller QDs, as discussed above, and the observation that the disparity between the data and the simple three level model (Equation (9)), is greatest for QD1. This trend of $k_0$ demonstrates the increased efficiency of thermally activated carrier trapping in QD monolayers as concentration is increased. This may arise as a consequence of intra-energy transfer; at the higher monolayer concentrations the QDs are sufficiently close to transfer energy to nearby QDs. A QD exciton can decay radiatively, have interactions with a trap, or transfer energy to excite a nearby QD. The exciton in the second QD can then also decay radiatively, interact with a trap, or transfer energy. Thus, effectively, the exciton created in the first QD has more opportunity to interact with traps. Reduced exciton peak emission has been reported in a number of high concentration QD systems.[41,47–51] Signatures of the intra energy transfer are masked in an ensemble measurement, however the possibility to interact with an increased number of trap states is expected to influence the PL decay of the ensemble. Other possible mechanisms could be direct tunnelling of carriers into trap states of neighbouring QDs or that closer packing of the QDs leads to the formation of additional defects. Further studies such as the temperature dependence of the spectrally filtered the emission would be required in order to more fully investigate the mechanism.

**Table 2**. Fitting parameters for Equation (14)

|  | $k_{Br}$ | $k_{Dx}$ | $\Delta E$ | $E_{LO}$ | $k_0$ | m | n |
|---|---|---|---|---|---|---|---|
| **QD1** |  |  |  |  | ± 9% | ± 0.05 | ± 0.05 |
| c1 | 0.128 ns$^{-1}$ | 0.019 ns$^{-1}$ | 7 meV | 20 meV | 0.03 ns$^{-1}$ | 2.2 | 1.85 |
| c2 | 0.175 ns$^{-1}$ | 0.020 ns$^{-1}$ | 7 meV | 20 meV | 0.06 ns$^{-1}$ | 2.2 | 1.85 |
| c3 | 0.208 ns$^{-1}$ | 0.022 ns$^{-1}$ | 7 meV | 20 meV | 0.09 ns$^{-1}$ | 2.2 | 1.85 |
| **QD2** |  |  |  |  | ± 10% | ±0.05 | ±0.05 |
| c1 | 0.068 ns$^{-1}$ | 0.025 ns$^{-1}$ | 4 meV | 20 meV | 0.001 ns$^{-1}$ | 2.8 | 2.6 |
| c2 | 0.070 ns$^{-1}$ | 0.026 ns$^{-1}$ | 4 meV | 20 meV | 0.009 ns$^{-1}$ | 2.8 | 2.6 |
| c3 | 0.076 ns$^{-1}$ | 0.032 ns$^{-1}$ | 4 meV | 20 meV | 0.03 ns$^{-1}$ | 2.8 | 2.6 |

**Conclusion**

The temperature dependence of the emission decay rates of CdTe QD monolayers with varying concentration have been investigated utilizing both steady-state PL and TRPL measurements. The steady-state PL properties can be explained by well-known equations and are used to extract the average phonon energy. The QD emission decay rate was shown to increase with temperature. The temperature dependence of the decay rate was strongly influenced by the concentration of QDs in the monolayer. The data at all concentrations was found to be well-explained with an analytic model involving bright and dark states, which can both interact with a trap state. Thermally activated carrier trapping had to be considered at temperatures > 150 K. Additionally, it has been shown that interactions with these trap states become more pronounced as the concentration of QDs is increased.

AUTHOR INFORMATION

**Corresponding Author**

Address correspondence to bradlel@tcd.ie Tel: +353-1-8963595

**Author Contributions**

The manuscript was written through contributions of all authors. All authors have given approval to the final version of the manuscript. ‡These authors contributed equally.

**Funding Sources**

This work was supported by Science Foundation Ireland (SFI) under grant number 10/IN.1/12975 and GPM acknowledges a postgraduate research scholarship from the Irish Research Council (RS/2011/287).

ACKNOWLEDGMENT

We would also like to acknowledge fruitful conversations with M.S Gaponenko.

Table of Contents Graphic

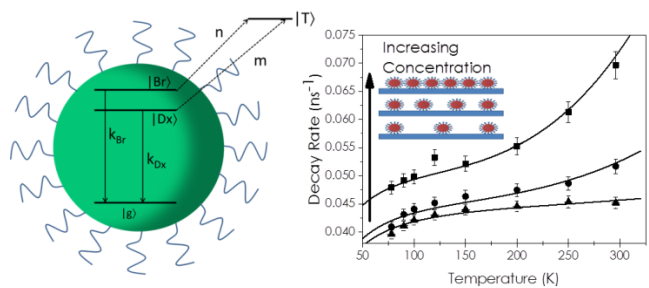